\documentclass[%
aip,
amsmath,amssymb,
reprint,%
]{revtex4-1}

\usepackage{amsmath,amssymb,bm}
\usepackage[overload]{empheq}
\usepackage{mathtools}
\usepackage{xcolor}
\usepackage[
	pdffitwindow=true,
	colorlinks=true,
	frenchlinks=false,
        linkcolor=blue,
	anchorcolor=blue,
        citecolor=blue,
        filecolor=blue,
        urlcolor=blue,
        bookmarks=true,
        bookmarksopen=true,
	bookmarksnumbered=true,
        bookmarksopenlevel=1,
        plainpages=false,
	pdfpagelayout=TwoPageLeft,
        pdfpagelabels=true,
	breaklinks
]{hyperref}
\usepackage[per-mode=symbol,separate-uncertainty]{siunitx}
\usepackage{graphicx}
\usepackage{dcolumn}
\usepackage{bm}
\usepackage[english]{babel}
\usepackage{color}
\usepackage{nicefrac}

\begin{document}
\preprint{AIP/123-QED}
\title[]{Anomalous spin Hall angle of a metallic ferromagnet determined by a multiterminal spin injection/detection device}

\author{T.~Wimmer}
\email[]{tobias.wimmer@wmi.badw.de}
\affiliation{Walther-Mei{\ss}ner-Institut, Bayerische Akademie der Wissenschaften, 85748 Garching, Germany}
\affiliation{Physik-Department, Technische Universit\"{a}t M\"{u}nchen, 85748 Garching, Germany}
\author{B.~Coester}
\affiliation{School of Physical and Mathematical Sciences, Nanyang Technological University, 21 Nanyang Link, Singapore 637371, Singapore}
\author{S.~Gepr{\"a}gs}
\affiliation{Walther-Mei{\ss}ner-Institut, Bayerische Akademie der Wissenschaften, 85748 Garching, Germany}
\author{R.~Gross}
\affiliation{Walther-Mei{\ss}ner-Institut, Bayerische Akademie der Wissenschaften, 85748 Garching, Germany}
\affiliation{Physik-Department, Technische Universit\"{a}t M\"{u}nchen, 85748 Garching, Germany}
\affiliation{Nanosystems Initiative Munich (NIM), Schellingstra{\ss}e 4, 80799 M\"{u}nchen, Germany}
\affiliation{Munich Center for Quantum Science and Technology (MCQST), Schellingstr. 4, D-80799 M\"{u}nchen, Germany}
\author{S.~T.~B.~Goennenwein}
\affiliation{Institut f{\"u}r Festk{\"o}rper- und Materialphysik and W{\"u}rzburg-Dresden Cluster of Excellence ct.qmat,
	Technische Universit{\"a}t Dresden, 01062 Dresden, Germany}
\author{H.~Huebl}
\affiliation{Walther-Mei{\ss}ner-Institut, Bayerische Akademie der Wissenschaften, 85748 Garching, Germany}
\affiliation{Physik-Department, Technische Universit\"{a}t M\"{u}nchen, 85748 Garching, Germany}
\affiliation{Nanosystems Initiative Munich (NIM), Schellingstra{\ss}e 4, 80799 M\"{u}nchen, Germany}
\affiliation{Munich Center for Quantum Science and Technology (MCQST), Schellingstr. 4, D-80799 M\"{u}nchen, Germany}
\author{M.~Althammer}
\email[]{matthias.althammer@wmi.badw.de}
\affiliation{Walther-Mei{\ss}ner-Institut, Bayerische Akademie der Wissenschaften, 85748 Garching, Germany}
\affiliation{Physik-Department, Technische Universit\"{a}t M\"{u}nchen, 85748 Garching, Germany}

\date{\today}


\begin{abstract}
	We report on the determination of the anomalous spin Hall angle in the ferromagnetic metal alloy cobalt-iron (Co$_{25}$Fe$_{75}$, CoFe). This is accomplished by measuring the spin injection/detection efficiency in a multiterminal device with nanowires of platinum (Pt) and CoFe deposited onto the magnetic insulator yttrium iron garnet (Y$_3$Fe$_5$O$_{12}$, YIG). Applying a spin-resistor model to our multiterminal spin transport data, we determine the magnon conductivity in YIG, the spin conductance at the YIG/CoFe interface and finally the anomalous spin Hall angle of CoFe as a function of its spin diffusion length in a single device. Our experiments clearly reveal a negative anomalous spin Hall angle of the ferromagnetic metal CoFe, but a vanishing ordinary spin Hall angle. This is in contrast to the results reported in Refs.~\onlinecite{Tian2016,Das2017} for the ferromagnetic metals Co and permalloy.
\end{abstract}

\maketitle

The spin Hall effect (SHE) is at the origin of a plethora of transport effects relevant for spintronics applications~\cite{Tian2016,Das2017,McGuire1975,HirschSHE,Maekawa2007,AltiSMR,Jungwirth2015,Klein2014,Wimmer2018}. While the charge to spin current conversion efficiency is conveniently expressed in terms of the phenomenological spin Hall angle $\Theta_\mathrm{SH}$, its microscopic origin is the spin-orbit interaction causing spin-selective scattering of charge carriers~\cite{Jungwirth2015,HoffmannSHE}. 
Many ferromagnetic metals exhibit a strong spin-orbit coupling, which manifests itself in various electrical transport effects, among them the anomalous Hall effect (AHE)~\cite{Ong2010}. The AHE hinges on the same physical principles as the SHE~\cite{McGuire1975,Jungwirth2015}. While the transverse charge current arising in the AHE has been studied for more than a century, the pure spin current part has only very recently received broad attention~\cite{Wang2014,Taniguchi2015,Tian2016,Das2017}.

Recent developments in magnetotransport experiments with incoherent magnons (the quantized excitations of the magnetization
)~\cite{CornelissenMMR,SchlitzMMR,KathrinLogik,Althammer2018} offer a suitable platform for the investigation of the SHE and the anomalous spin Hall effect (ASHE) in ferromagnets~\cite{Tian2016,Das2017,Das2018,Gibbons2018,Iihama2018}. In these experiments, a spin current is injected into an adjacent magnetic insulator via the SHE. More specifivally, a DC charge current in a metallic electrode generates a spin accumulation at the interface between the metal and the magnet, which in turn induces a non-equilibrium magnon accumulation in the magnet. The non-equilibrium magnons diffuse in the magnet, and are detected in a second, electrically separate metallic electrode as a voltage signal via the inverse SHE. In a recent work, Das et al.~reported spin injection and detection in YIG via the AHE~\cite{Ong2010} using Py electrodes~\cite{Das2017}. They found a magnetic field enhanced injection/detection efficiency of the permalloy (Py) electrodes due to the gradual increase of the effective anomalous spin Hall angle $\Theta_\mathrm{ASH}$ in Py. 

\begin{figure}[]%
	\includegraphics[]{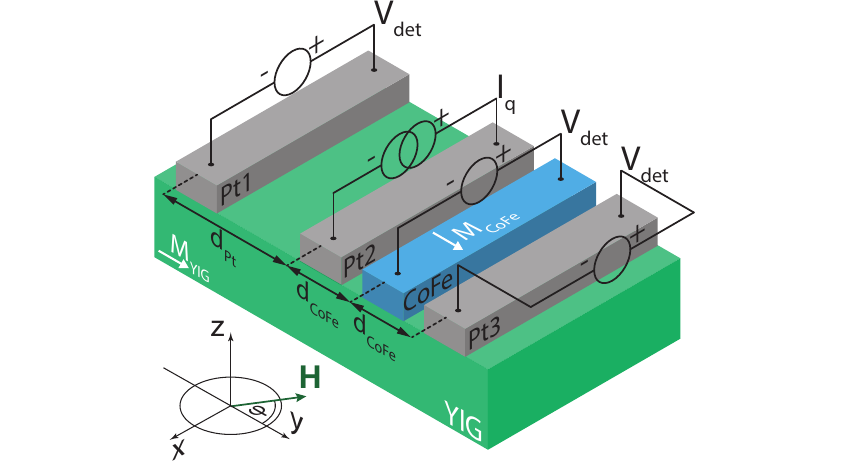}%
	\caption{Schematic depiction of the device, the electrical connection scheme and the coordinate system. A charge current $I_\mathrm{q}$ is fed through the Pt2 electrode, resulting in a spin current injection into YIG via the SHE. The lateral diffusion of the magnon spin current is electrically detected at the Pt electrodes ('Pt1' and 'Pt3') and the ferromagnetic metal electrode ('CoFe') as the detector voltage $V_\mathrm{det}$. The center-to-center distances between each of the Pt electrodes is constant, such that $d_\mathrm{Pt}=2d_\mathrm{CoFe}$. 
	}%
	\label{fig:scheme}%
\end{figure}

In this Letter, we report on the determination of $\Theta_\mathrm{ASH}$ of the ferromagnetic metal alloy Co$_{25}$Fe$_{75}$ (CoFe)~\cite{Schoen2016} via all-electrical magnon transport measurements in the magnetic insulator YIG. For this purpose, we utilize a multiterminal structure with four metallic electrodes -- one made of CoFe and three made of Pt -- deposited onto a YIG thin film (see Fig.~\ref{fig:scheme}).


Our device consists of a $\SI{1}{\micro\metre}$ thick, commercially availabe YIG film grown on a GGG (Gd$_{3}$Ga$_{5}$O$_{12}$) substrate via liquid phase epitaxy. Both the Pt and the CoFe electrodes were deposited by DC sputtering and patterned via electron beam lithography and lift-off~\cite{Wimmer2018}. The CoFe electrode was additionally capped with a $\SI{2.5}{\nano\metre}$ thick Al layer to prevent oxidation. In a further step, Al leads and bondpads were deposited to connect the device electrically. Each electrode has a width of $w=\SI{500}{\nano\metre}$ and a thickness of $t_\mathrm{Pt}=t_\mathrm{CoFe}=\SI{7}{\nano\metre}$. The lengths of the strips are $l_\mathrm{Pt1}=l_\mathrm{Pt3} = \SI{148}{\micro\metre}$ for the outer electrodes and $l_\mathrm{Pt2}=l_\mathrm{CoFe} = \SI{162}{\micro\metre}$ for the inner ones. As indicated in Fig.~\ref{fig:scheme}, the center-to-center distances between the metal strips are $d_\mathrm{Pt}=\SI{1.6}{\micro\metre}$ 
and $d_\mathrm{CoFe}=\SI{0.8}{\micro\metre}$ (cf.~Fig.~\ref{fig:scheme}). 
For the injection of magnons, we apply a charge current $I_\mathrm{q}=\SI{0.5}{\milli\ampere}$ to the Pt2 electrode (the injector) and detect the magnon transport signal as the detector voltage $V_\mathrm{det}$ at the Pt1, Pt3 and CoFe electrodes (see Fig.~\ref{fig:scheme}). 
In order to distinguish between electrically (via the SHE) and thermally (via Joule heating) injected magnons, we utilize the current reversal method~\cite{SchlitzMMR,KathrinLogik}. Here, we focus on the magnon transport via the electrical SHE-induced spin current injection. All measurements are conducted in a superconducting magnet cryostat at a constant temperature of $T=\SI{280}{\kelvin}$~\footnote{This specific temperature was used due to better temperature stability in our cryostat}. 
In order to compare between different detector signals, we define a normalized signal amplitude as $R_\mathrm{det} = ({V_\mathrm{det}/I_\mathrm{q}})\cdot({A_\mathrm{inj}/A_\mathrm{det}})$, which 
accounts for the different interface areas $A_\mathrm{inj}$ ($A_\mathrm{det}$) of the injector (detectors)~\cite{KathrinLogik}. 

\begin{figure}[]%
	\includegraphics[]{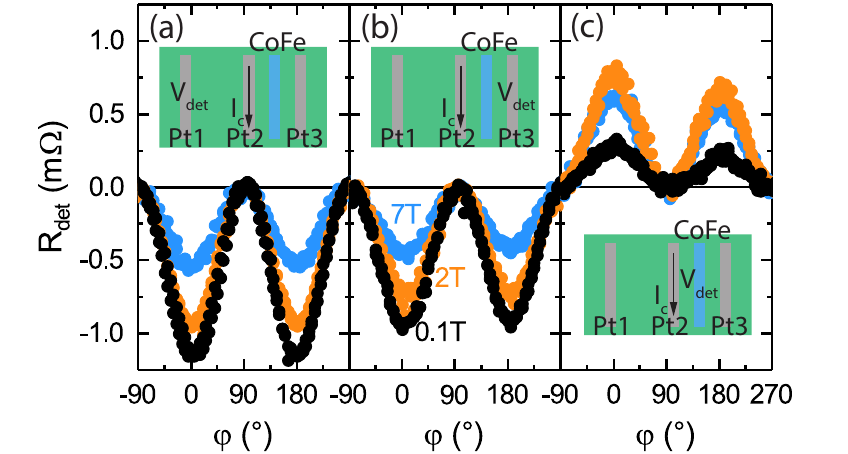}%
	\caption{Resistance $R_\mathrm{det}$ measured using different detector electrodes. (a) The signal at the Pt1 detector is taken as a reference measurement, with which we can characterize the magnon transport in the YIG layer. 
	(b) The Pt3 detector signal is somewhat smaller than the Pt1 signal owing to the finite absorption of the magnon current by the CoFe electrode in between Pt2 and Pt3. (c) The $R_\mathrm{det}$ associated with the detector voltage recorded at the CoFe electrode shows a sign reversal, indicating that the anomalous spin Hall angle in CoFe is negative. 
	}%
	\label{fig:rotations}%
\end{figure}

To characterize the magnon transport in our device, we measure $R_\mathrm{det}$ as a function of the magnetic field orientation $\varphi$ for various in-plane field magnitudes $\mu_0 H$. 
Corresponding data are shown in Fig.~\ref{fig:rotations} (a)-(c) for three different external magnetic fields. Firstly, panel (a) shows the reference measurement using Pt1 as a detector. In accordance with Refs.~\cite{CornelissenMMR,SchlitzMMR,CornelissenField} 
, we observe a 
$\sin^2(\varphi)$-dependence of $R_\mathrm{det}$ with reduced amplitudes for increasing external magnetic field strengths. 
Secondly, panel (b) shows $R_\mathrm{det}$ recorded across Pt3. Since the separation of the Pt1 and Pt3 strip to the injector strip Pt2 are the same (cf.~\ref{fig:scheme}), one would expect the same signal magnitude. 
However, the $R_\mathrm{det}$ modulation recorded across the Pt3 strip is significantly smaller, which we attribute to a partial absorption of the magnon spin current in the CoFe electrode located in between the Pt2 and Pt3 electrodes. Finally, panel (c) shows $R_\mathrm{det}$ measured at the CoFe electrode. Interestingly, the polarity of the detected voltage is inverted. Since all strips were contacted with identical polarity in the experiments (see Fig.~\ref{fig:scheme}), we conclude that the anomalous spin Hall angle $\Theta_\mathrm{ASH}^\mathrm{CoFe}$ in CoFe is negative compared to the positive spin Hall angle $\Theta_\mathrm{SH}^\mathrm{Pt}$ in Pt~\cite{Weiler2013,HoffmannSHE}. This is in agreement with the negative spin Hall angles reported for both Co and Fe~\cite{Wang2014}. Unlike the magnetic field suppression observed for the Pt detector strips, we find a significant enhancement of $R_\mathrm{det}$ for increasing magnetic fields up to $\mu_0 H = \SI{2}{\tesla}$ for the CoFe detector. We attribute this to the field-induced increase of $\Theta_\mathrm{ASH}^\mathrm{CoFe}$, qualitatively similar to the results reported in Ref.~\cite{Das2017}. For larger magnetic fields ($\mu_0 H = \SI{7}{\tesla}$), however, we observe a suppression of the magnon transport signal. Since the CoFe magnetization $M_\mathrm{CoFe}$ saturates around $\mu_0 H = \SI{2}{\tesla}$~\cite{Schoen2016}, we attribute this field suppression to the YIG magnon system in analogy to the situation observed for the Pt detectors~\cite{CornelissenMMR}. 
Interestingly, we observe a distinct asymmetry in the magnitudes of the signal for strong magnetic fields $\mu_0 H > \SI{2}{\tesla}$, which is discussed in more detail in the Supplementary Information (SI)~\footnote{\label{fn:supp}See Supplemental Information, where we present additional measurements of the spin Hall magnetoresistance of the YIG/Pt interface, the magnon diffusion length and magnon conductivity of the YIG film, a more detailed study of the spin conductance of the YIG/CoFe interface, a study on the influence of the shape anisotropy of the CoFe electrode and a more in depth investigation of the asymmetry feature observed in the magnon transport signals using the CoFe electrodes.}.

\begin{figure}[]%
	\includegraphics[]{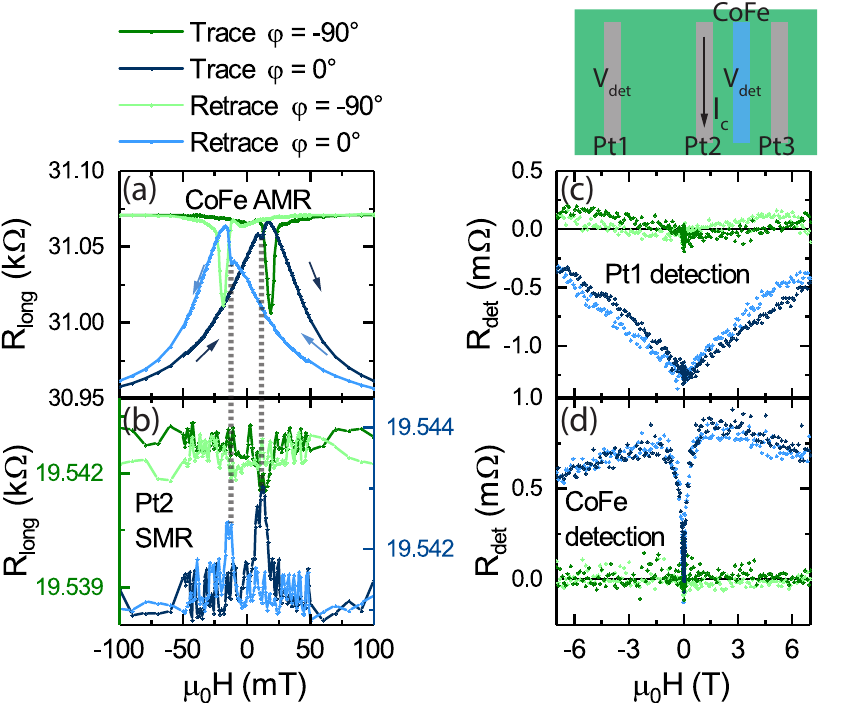}%
	\caption{Longitudinal resistance $R_\mathrm{long}$ and magnon transport signal $R_\mathrm{det}$ measured as a function of the magnetic field strength for field directions pointing perpendicular (blue) and parallel (green) to the strip length. Dark and light colored lines correspond to up- and down-sweep curves, respectively. (a) $R_\mathrm{long}$ measured on the CoFe electrode, showing the AMR  
		with a switching field of $\mu_0 H = \pm\SI{18}{\milli\tesla}$. Additionally, a second switching at lower fields $\mu_0 H = \pm\SI{11}{\milli\tesla}$ (indicated by gray dashed lines) is observed, which corresponds to the $R_\mathrm{long}$ change measured on the Pt2 electrode (SMR) in (b). Here, the green curves correspond to the left vertical axis, while the blue lines refer to the right axis. (c), (d) show $R_\mathrm{det}$ as a function of the magnetic field strength measured at the Pt1 and CoFe detector, respectively. 
	}%
	\label{fig:sweeps}%
\end{figure}

In Fig.~\ref{fig:sweeps} (a), we plot the anisotropic magnetoresistance (AMR) of the CoFe electrode by measuring its longitudinal resistance $R_\mathrm{long}$ as a function of the magnetic field strength (the sweep direction is indicated by arrows). The blue (green) colored lines correspond to the field direction pointing perpendicular (parallel) to the strip length, while dark (light) colored lines correspond to the up (down) sweep of the magnetic field strength (trace and retrace, respectively). Obviously, we observe a clear AMR with a maximum (minimum) in resistance for parallel (perpendicular) field alignment with respect to the strip length for a coercive field of approximately $\pm\SI{18}{mT}$. Additionally, we find a second peak at a characteristic field of roughly $\pm\SI{11}{\milli\tesla}$. This feature corresponds well to the switching field observed for the longitudinal resistance of the Pt2 electrode, which is shown in Fig.~\ref{fig:sweeps} (b) (switching field indicated by gray dashed lines). Therefore, the peaks in the resistance of the CoFe strip around $\SI{11}{\milli\tesla}$ can be attributed to a spin Hall magnetoresistance (SMR)~\cite{AltiSMR} contribution related to the magnetization reversal in the YIG film. Most importantly, these magnetoresistance measurements show that exchange coupling between the two ferromagnetic layers is not relevant, since no exchange bias effect can be observed (which would lead to a shift of the hysteresis curves along the magnetic field axis). Figure~\ref{fig:sweeps} (c), (d) show the detector signals $R_\mathrm{det}$ as a function of the magnetic field strength measured at the Pt1 and CoFe electrodes, respectively. For the reference detector (Pt1), we find a continuous suppression of the magnon transport signal $R_\mathrm{det}$ with increasing magnetic field strength when the field is oriented perpendicular to the strips (blue data points in Fig.~\ref{fig:sweeps} (c)). For a parallel alignment of the field and the strips (green data points), the signal vanishes. This response is quantitatively consistent with the field-orientation dependent data (Fig.~\ref{fig:rotations}). Figure~\ref{fig:sweeps} (d) shows the magnetic field dependence of $R_\mathrm{det}$ when the CoFe electrode is used as the detector. Here, $R_\mathrm{det}$ is zero for $\mu_0 H = 0$ for both field orientations. When the field is oriented prependicular to the strips, $R_\mathrm{det}$ rapidly increases 
and reaches its maximum at around $\mu_0 H \approx \SI{2}{\tesla}$. Since the injection and detection efficiency of the magnons is maximized when the magnetizations $M_\mathrm{YIG}$ and $M_\mathrm{CoFe}$ are aligned perpendicular to the electrodes~\cite{Das2017}, the maximum of $R_\mathrm{det}$ is expected when 
the magnetization $M_\mathrm{CoFe}$ is fully saturated perpendicularly to the strips overcoming the shape anisotropy at around $\SI{2}{\tesla}$~\cite{Das2017}. For larger magnetic fields, we again observe a field-induced suppression of the signal, which was already discussed for the orientation dependent measurements in Fig.~\ref{fig:rotations} and originates from the field dependence of the magnon transport in YIG. Following Ref.~\cite{Das2017}, the contributions of $\Theta_\mathrm{SH}$ and $\Theta_\mathrm{ASH}$ can be separated by identifying the magnon transport signal at the switching field and the saturation field of the CoFe detector, respectively. It is, however, evident that the CoFe detector signal in Fig.~\ref{fig:sweeps} (d) becomes zero for small magnetic fields, suggesting that there is no contribution from a pure SHE in CoFe, in contrast to the results reported for Py~\cite{Das2017}. We note, however, that this observation strongly depends on whether or not the CoFe is in a multidomain state, since a net magnetization $M_\mathrm{CoFe}$ perpendicular to the electrode could be counterbalanced by a positive SHE contribution. Clearly, we can verify again the asymmetry feature when the CoFe electrode is used as a detector~\footnotemark[2].

\begin{figure}[]%
	\includegraphics[]{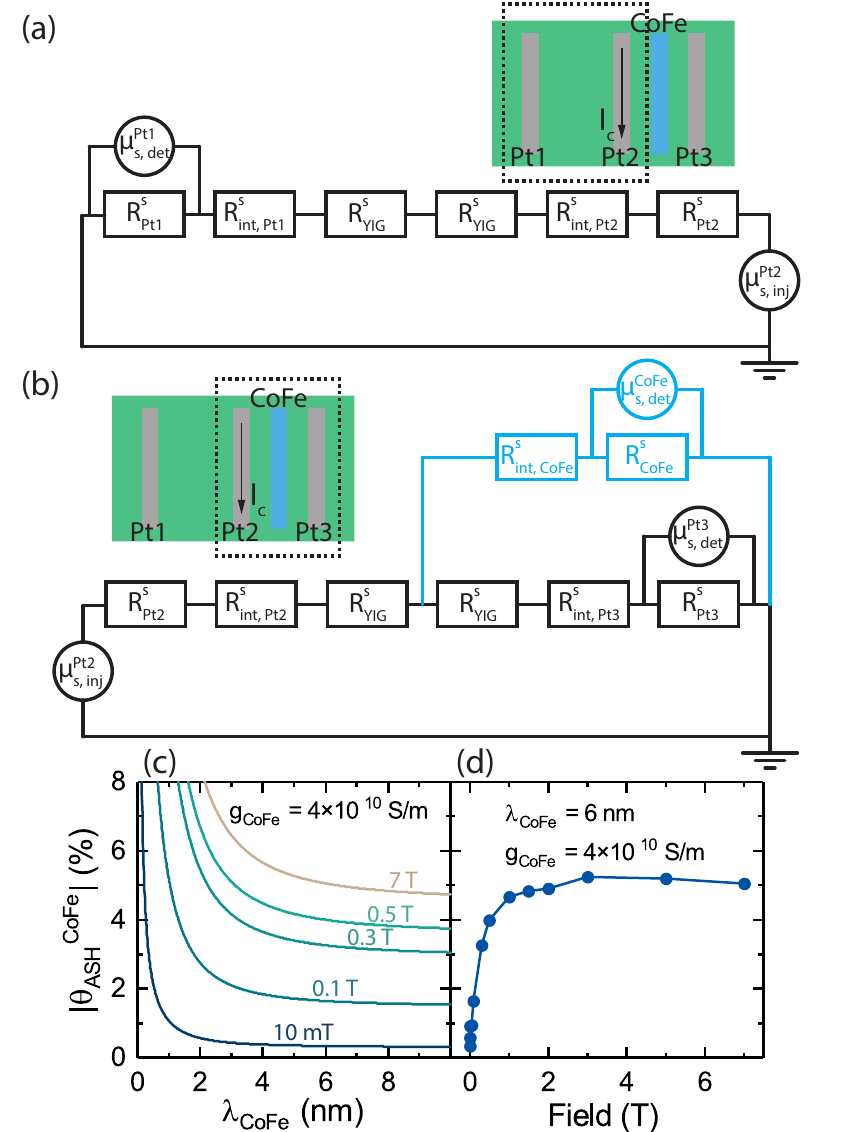}%
	\caption{Equivalent spin-resistor network for the Pt2-Pt1 contact pair (a) and the Pt2-CoFe-Pt3 contact configuration (b).~(c) Experimentally determined absolute value of the anomalous spin Hall angle $\Theta_\mathrm{ASH}^\mathrm{CoFe}$ as a function of the spin diffusion length $\lambda_\mathrm{CoFe}$ for various external magnetic fields. Here, the spin conductance of the YIG/CoFe interface was set to a constant value $g_\mathrm{CoFe} = \SI{4e10}{\siemens\per\metre}$. (d) Anomalous spin Hall angle of CoFe as a function of the applied magnetic fields, assuming a spin diffusion length $\lambda_\mathrm{CoFe} = \SI{6}{\nano\metre}$. 
	}%
	\label{fig:theta}%
\end{figure}

Utilizing our multiterminal magnon transport device, we are able to extract the anomalous spin Hall angle $\Theta_\mathrm{ASH}^\mathrm{CoFe}$. To this end, we model the spin transport in our device by employing the spin-resistor circuit model proposed in Ref.~\cite{CornelissenTheory}. This approach is valid as long as the distance $d$ between the considered electrodes is smaller than the characteristic magnon diffusion length $\lambda_\mathrm{m}$ in our YIG film. We verify that $d_\mathrm{Pt}=2d_\mathrm{CoFe}<\lambda_\mathrm{m}\approx\SI{6}{\micro\metre}$ for a comparable YIG film (see SI). The equivalent spin-resistor circuit diagram for the Pt2-Pt1 contact pair and the three-terminal Pt2-CoFe-Pt3 contacts are shown in Fig.~\ref{fig:theta} (a) and (b), respectively. Here, the individual resistors are described by three different resistances: firstly, $R_{i}^\mathrm{s} = \lambda_{i}\rho_{i}/[l_{i} w \tanh(t_{i}/\lambda_{i})]$ is the spin resistance of electrode $i$ (with $i$ = Pt1, CoFe, Pt3) with $\lambda_{i}$ the spin diffusion length and $\rho_i$ the electrical resistivity. 
Furthermore, $t_i$, $l_i$ and $w$ denote the thickness, length and width of electrode $i$. Secondly, $R_{\mathrm{int,} i}^\mathrm{s} = 1/(g_\mathrm{i}l_{i}w)$ is the interface spin resistance, with $g_i$ the interface spin conductance of electrode $i$ and lastly $R_\mathrm{YIG}^\mathrm{s} = d_\mathrm{CoFe}/(\sigma_\mathrm{m}l_\mathrm{Pt2}t_\mathrm{YIG})$ is the YIG spin resistance for a distance of $d_\mathrm{CoFe}$ with $\sigma_\mathrm{m}$ the magnon conductivity and $t_\mathrm{YIG}$ the thickness of the YIG film. 
For both circuits shown in Fig.~\ref{fig:theta} (a) and (b), the "spin battery" of the network is characterized by the injected spin chemical potential $\mu_\mathrm{s, inj}^\mathrm{Pt2} = 2\Theta_\mathrm{SH}^\mathrm{Pt} I_\mathrm{c}  \lambda_\mathrm{Pt} [{R_\mathrm{Pt2}}/{l_\mathrm{Pt2}}] \tanh\left({t_\mathrm{Pt}}/({2\lambda_\mathrm{Pt}})\right)$~\cite{ChenSMR} at the YIG/Pt2 interface. Here, $\lambda_\mathrm{Pt}$ is the spin diffusion length of Pt, $\Theta_\mathrm{SH}^\mathrm{Pt}$ is the spin Hall angle of Pt and $R_\mathrm{Pt2}$ is the electrical resistance of the Pt2 electrode. The spin chemical potential "drop" across each detector $i$ is given via the measured detector voltages $V_\mathrm{det}^{i}$ as $ \mu_\mathrm{s, det}^{i} = {2t_\mathrm{i}}/({\Theta_\mathrm{(A)SH}^\mathrm{i}l_\mathrm{i}}) 
\left(1+ [{\cosh(t_{i}/\lambda_{i})-1}]^{-1}\right) V_\mathrm{det}^{i}$~\cite{CornelissenTheory}.~For each detector $i$, we can then calculate the spin transfer efficiency as
\begin{equation}
\eta_\mathrm{s}^{i}=\frac{\mu_\mathrm{s, det}^{i} }{\mu_\mathrm{s, inj}^\mathrm{Pt2}}.
\label{eq:eta}
\end{equation}
Applying Kirchhoff's laws to the spin-resistor network shown in Fig.~\ref{fig:theta} (a), we obtain the spin transfer efficiency of the Pt1 detector as
\begin{equation}
	\eta_\mathrm{s}^\mathrm{Pt1} = \frac{R_\mathrm{Pt1}^\mathrm{s}}{R_\mathrm{Pt2}^\mathrm{s}+R_\mathrm{Pt1}^\mathrm{s}+R_\mathrm{int, Pt2}^\mathrm{s}+R_\mathrm{int, Pt1}^\mathrm{s}+2R_\mathrm{YIG}^\mathrm{s}},
\label{eq:etaPt}
\end{equation}
while analyzing the circuit shown in Fig.~\ref{fig:theta} (b), we find
\begin{subequations}
	\begin{align}
	\eta_\mathrm{s}^\mathrm{Pt3} = \frac{R_\mathrm{Pt3}^\mathrm{s}\zeta}{R_\mathrm{tot}^\mathrm{s} \left(1 + \zeta\right)},  \label{eq:etaPt3} \\
	\eta_\mathrm{s}^\mathrm{CoFe} = \frac{R_\mathrm{CoFe}^\mathrm{s}}{R_\mathrm{tot}^\mathrm{s}\left(1+\zeta\right)}, \label{eq:etaCoFe}
	\end{align}
\end{subequations}
for the Pt3 and CoFe detectors. Here, $\zeta = [R_\mathrm{int, CoFe}^\mathrm{s}+R_\mathrm{CoFe}^\mathrm{s}]/[R_\mathrm{YIG}^\mathrm{s}+R_\mathrm{int, Pt3}^\mathrm{s}+R_\mathrm{Pt3}^\mathrm{s}]$ and $R_\mathrm{tot}^\mathrm{s}$ is the total resistance of the spin-resistor network of Fig.~\ref{fig:theta} (b).

On the basis of this model, we now calculate $\sigma_\mathrm{m}$, $g_\mathrm{CoFe}$ of the YIG/CoFe interface and finally $\Theta_\mathrm{ASH}^\mathrm{CoFe}$ of CoFe. We obtain $\sigma_\mathrm{m}$ by equating Eqs.~\eqref{eq:eta} and~\eqref{eq:etaPt} for $i=\mathrm{Pt1}$~\footnotemark[2]. Note, that the spin conductance $g_\mathrm{Pt}$ for the YIG/Pt interfaces was independently determined via longitudinal SMR measurements~\footnotemark[2]. We extract $\sigma_\mathrm{m}$ for the different magnetic field values and find $\sigma_\mathrm{m}=\SI{3.2e4}{\siemens\per\metre}$ for an external magnetic field of $\mu_0 H = \SI{0.1}{\tesla}$~\cite{}. 
In a next step, we extract $g_\mathrm{CoFe}$. Since our experiment does not allow to determine the spin diffusion length $\lambda_\mathrm{CoFe}$ of CoFe, we determine $g_\mathrm{CoFe}$ as a function of $\lambda_\mathrm{CoFe}$ from Eqs.~\eqref{eq:eta} and~\eqref{eq:etaPt3} for $i=\mathrm{Pt3}$. Substituting the values extracted for $\sigma_\mathrm{m}$ for each of the magnetic fields measured, we find that $g_\mathrm{CoFe}$ only varies by $\sim\SI{0.05}{\percent}$ when changing $\lambda_\mathrm{CoFe}$ from $\SI{0}{\nano\metre}$ to $\SI{10}{\nano\metre}$. 
Moreover, we find $g_\mathrm{CoFe}$ to vary between approximately $\SI{2e10}{\siemens\per\metre}$ for $\mu_0 H =\SI{7}{\tesla}$ and $\SI{4e10}{\siemens\per\metre}$ for $\mu_0 H =\SI{0.5}{\tesla}$. Since the spin conductance is not expected to depend on the applied magnetic field, we adopt a constant value of $g_\mathrm{CoFe} = \SI{4e10}{\siemens\per\metre}$ in the following~\footnotemark[2]. 
Then, we can extract $\Theta_\mathrm{ASH}^\mathrm{CoFe}$ as a function of $\lambda_\mathrm{CoFe}$ from Eqs.~\eqref{eq:eta} and~\eqref{eq:etaCoFe} for $i=\mathrm{CoFe}$. The result is shown in Fig.\ref{fig:theta} (c) for different magnetic fields. Obviously, $\Theta_\mathrm{ASH}^\mathrm{CoFe}$ saturates as a function of $\lambda_\mathrm{CoFe}$ at around $\sim\SI{7}{\nano\metre}$, which corresponds to the CoFe electrode thickness $t_\mathrm{CoFe}$. This is reasonable, since (experimentally) we do not expect any change of $\Theta_\mathrm{ASH}^\mathrm{CoFe}$ for $\lambda_\mathrm{CoFe}>t_\mathrm{CoFe}$. Finally, we estimate the field dependence of $\Theta_\mathrm{ASH}^\mathrm{CoFe}$ by assuming $\lambda_\mathrm{CoFe}=\SI{6}{\nano\metre}$~\cite{Zahnd2018}. Note, that the value of $\lambda_\mathrm{CoFe}$ only affects the quantitative values for $\Theta^\mathrm{CoFe}_\mathrm{ASH}$, but the qualitative field dependence remains the same. 
Plotting $\Theta_\mathrm{ASH}^\mathrm{CoFe}$ as a function of magnetic field in Fig.~\ref{fig:theta} (d), we find that $\Theta_\mathrm{ASH}^\mathrm{CoFe}$ rapidly increases with increasing magnetic field (as $M_\mathrm{CoFe}$ saturates) and reaches its maximum value for about \SI{2}{\tesla}-\SI{3}{\tesla} at $\sim\SI{5}{\percent}$. For permalloy, a spin Hall angle of $\SI{2}{\percent}$ was reported~\cite{Wang2014}. Clearly, the field dependence of $\Theta_\mathrm{ASH}^\mathrm{CoFe}$ in our experiment is determined by the magnetization $M_\mathrm{CoFe}$ aligning perpendicularly to the CoFe strip length (i.e.~along the magnetic hard axis). As detailed in the SI, however, the application of a Stoner-Wohlfarth model with uniaxial shape anisotropy~\cite{Stoner1948} does not reproduce the observed field dependence well, suggesting that the CoFe electrode is in a multidomain state for small magnetic fields. 

In conclusion, we demonstrated the determination of the anomalous spin Hall angle $\Theta_\mathrm{ASH}^\mathrm{CoFe}$ of the ferromagnetic metal Co$_{25}$Fe$_{75}$ employing a multiterminal spin injection/detection device. Using both paramagnetic Pt and ferromagnetic CoFe electrodes on the ferrimagnetic insulator YIG, we were able to determine the magnon conductivity of YIG, the spin conductance of the YIG/CoFe interface and finally the anomalous spin Hall angle of CoFe on a single device. We based our analysis on a spin-resistor model~\cite{CornelissenTheory} and found that the pure SHE contribution in CoFe is negligible, which is in contrast to the finite SHE contribution reported for Py~\cite{Das2017}. The anomalous spin Hall angle of CoFe was found to increase strongly by saturating $M_\mathrm{CoFe}$ with an applied magnetic field and shows a saturation value of $\sim\SI{5}{\percent}$ for magnetic fields of $\mu_0 H \gtrsim \SI{2}{\tesla}$.

This work is funded by the Deutsche Forschungsgemeinschaft (DFG, German Research Foundation) under Germany’s Excellence Strategy -- EXC-2111 -- 390814868 and project AL2110/2-1.

%

\end{document}